\documentclass{JAC2003}

\usepackage{amsmath}
\usepackage{graphicx}
\usepackage{booktabs}
\usepackage{epstopdf}
\usepackage{placeins}
\usepackage{hyperref}
\usepackage[usenames,dvipsnames]{color}
\usepackage{comment}

\definecolor{darkred}{rgb}{0.5,0,0}
\definecolor{darkgreen}{rgb}{0,0.5,0}
\definecolor{darkblue}{rgb}{0,0,0.5}
\hypersetup{ colorlinks,
linkcolor=darkred,
filecolor=darkgreen,
urlcolor=darkblue,
citecolor=darkred }

\begin{document}
\title{OPERATIONAL BEAMS FOR THE LHC}

\author{Y. Papaphilippou, H. Bartosik, G. Rumolo, D. Manglunki, CERN, Geneva, Switzerland}

\maketitle

\begin{abstract}
The variety of beams, needed to set-up in the injectors as requested in
the LHC, are reviewed, in terms of priority but also performance expectations and
reach during 2015. This includes the single bunch beams for machine commissioning and measurements (probe, Indiv) but also the standard physics beams with 50 ns and 25 ns bunch spacing and their high brightness variants using the Bunch Compression Merging and Splitting (BCMS) scheme. The required parameters and target performance of special beams like the doublet for electron cloud enhancement and the more exotic 8b$\oplus$4e beam, compatible
with some post-scrubbing scenarios are also described. The progress and
plans for the LHC ion production beams during 2014-2015 are detailed. Highlights on the current progress of the setting up of the various beams are finally presented with special emphasis on potential performance issues across the proton and ion injector chain.
\end{abstract}

\section{INTRODUCTION}

During the LHC Run 1, the LHC physics production was based on beams with 50~ns bunch spacing. Beams with 25~ns bunch spacing were injected into LHC on few occasions for injection tests, Machine Developments (MDs), the scrubbing run followed by a pilot physics run~\cite{Iadarola:Evian2014}. After the startup in 2015, apart from the LHC collision energy which will be raised to 6.5~TeV per beam, it will be crucial to establish physics operation with the nominal 25~ns bunch spacing in order to maximise the integrated luminosity in Run 2 for the limited event pile-up acceptable by the LHC experiments~\cite{Gorini:Evian2014}. The LHC will thus request a large variety of beams, including single bunches for machine commissioning and measurements but also the standard physics beams with 50~ns and 25~ns bunch spacing and their high brightness variants using the Bunch Compression Merging and Splitting (BCMS) scheme~\cite{Garoby:BCMS,Damerau:IPAC13}. In addition, special beams like the doublet for electron cloud enhancement \cite{Iadarola:Evian2014} and the more exotic 8b$\oplus$4e beam~\cite{Damerau:RLIUP}, compatible with some post-scrubbing scenarios should be also prepared and made available from the injectors.

This paper reviews the parameters of the LHC physics beams achieved in the injectors until 2012 and the expectations for their performance in the following run (for a detailed analysis see~\cite{Bartosik:Evian2014}). The progress and
plans for the LHC ion production beams during 2014-2015 are also finally presented.

\section{SINGLE BUNCH BEAMS}

During the preparation of the  LHC p-Pb run in 2013, a new improved  production scheme has been developed \cite{Hancock:Indivs}, with which single bunch LHC beams can be generated in the PSB. The main ingredient
 was the revision of the controlled longitudinal blow up during first part of PSB cycle, through optimisation of C16 and C02 parameters. Thereby, the C16 voltage can be used
 for intensity control. This assures the preservation of the 6D phase space volume for different intensities with excellent shot-to-shot reproducibility and control of both intensity and
longitudinal emittance.  
It is therefore expected that after Long Shutdown 1 (LS1) the injectors will be able to deliver LHCPROBE bunches ($5\!\times\!10^9 - 2\!\times\!10^{10}$\,p/b) and LHCINDIV bunches ($2\!\times\!10^{10} - 3\!\times\!10^{11}$\,p/b) to the LHC with smaller intensity fluctuations compared to the operation during Run 1. The LHCINDIV parameter range was also extended in MDs to produce single bunches with up to $4\!\times\!10^{11}$\,p/b and/or with lower longitudinal emittances (down to 0.15 eVs), at SPS injection. These high intensity variants can be used for impedance or beam-beam studies. Finally, a procedure for producing
Gaussian bunches for Van der Meer scans was established in 2012. It is based on longitudinal and transverse shaving in the PSB to obtain ÒlargeÓ emittance (more than 2.5~$\mu$m) single bunches with under-populated tails. Because of diffusion processes in the PS and SPS, these bunches evolve into almost perfect Gaussian shapes at the exit of the SPS and at collision in the LHC as confirmed by the experiments. This beam will need to be ready for the van der Meer scans at the beginning of the 2015 run and can profit from the newly established single bunch production scheme in the PSB.

\section{LHC PHYSICS BEAMS}

LHC operation during Run 1 used mainly 50~ns beams produced with the standard scheme of bunch splittings in the PS. Beams with the nominal 25~ns bunch spacing have been used in the LHC for the scrubbing run and machine development studies. With the successful implementation of the BCMS scheme~\cite{Garoby:BCMS,Damerau:IPAC13} in the PS in 2012, the injectors were also able to provide LHC beams with almost twice the brightness compared to the standard production schemes. While the 50~ns BCMS beam was injected into the LHC only an emittance preservation study of a high brightness beam along the LHC ramp, the 25~ns BCMS beam was used for the 25~ns pilot physics run at the end of 2012. 
\begin{figure}[b]
    \centering
     \includegraphics[trim= 0 0 0 0, clip,width=0.49\textwidth]{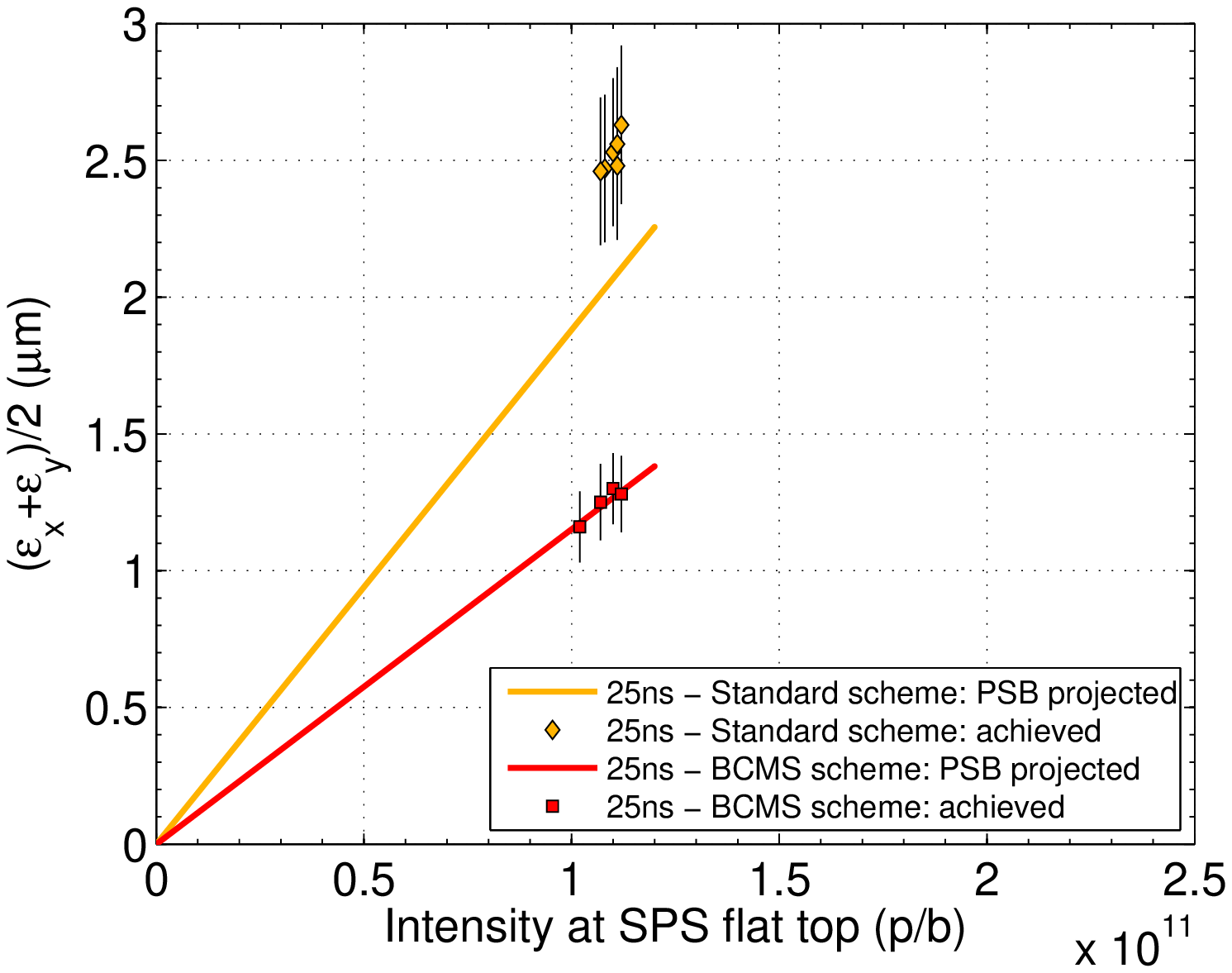}
     \includegraphics[trim= 0 0 0 0, clip,width=0.49\textwidth]{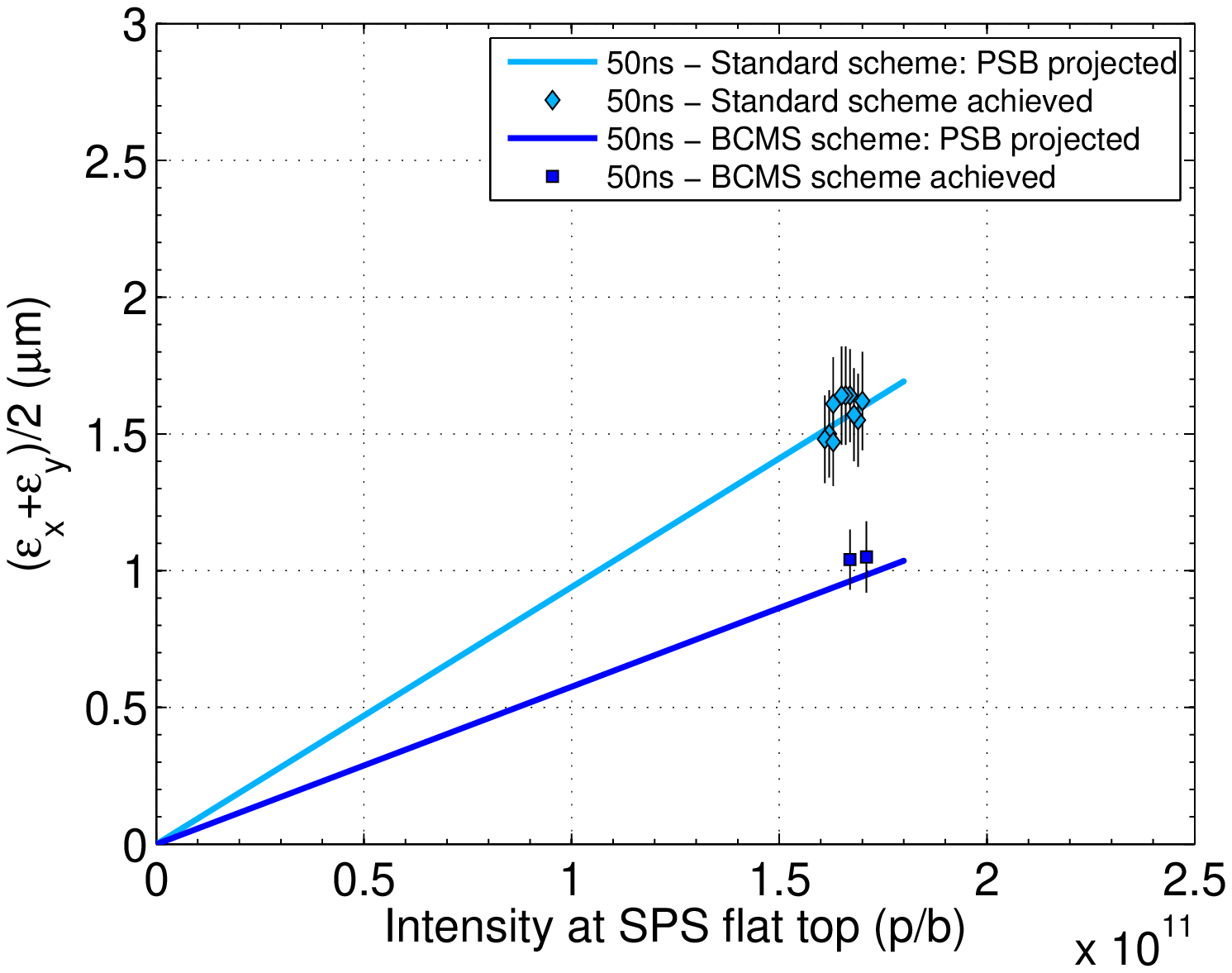}    
    \caption{Beam parameters achieved operationally in the SPS in 2012 with the Q20 optics for 50 ns beams (bottom) and 25 ns beams (top) extracted to the LHC.}
    \label{FIG:2012Performance}
\end{figure}
It should be emphasised that all these LHC beams were produced close to the performance limits of the injector chain: For the 50~ns beam
the intensity per bunch is close to the limit of longitudinal
instability in the PS, whereas the brightness of the BCMS beam is at the present space charge limit in the SPS. For the 25~ns beam, the intensity per bunch is close to the limit of RF power
and longitudinal instability in the SPS while the brightness is at the present space charge limit in the PS. Figure~\ref{FIG:2012Performance} shows the beam parameters for the two types of beams as achieved in 2012 after the operational deployment of the Q20 low gamma transition optics in the SPS \cite{Bartosik:Q20, Papaphilippou:Q20}. The transverse emittances shown in these plots are deduced from combined wire-scans at the end of the SPS flat bottom and the values were cross-checked with measurements in the LHC. The error bars include the spread over several measurements as well as a systematic uncertainty of 10\%. The bunch intensity is measured at the SPS flat top after the scraping of the beam tails, as required prior to extraction into LHC. The solid lines correspond to the PSB brightness curve (i.e.~the emittance as a function of intensity measured at PSB extraction) translated into protons per SPS bunch for each beam type assuming intensity loss and emittance growth budgets of 5~\% in the PS and 10~\% in the SPS, respectively. All beams were produced within the allocated budgets for beam degradation along the injector chain apart from the standard 25~ns beam, which suffers from slow losses at the SPS flat bottom and maybe also from space charge effects at the PS injection. Nevertheless, the nominal 25\,ns beam is well within the original
specifications (i.e.~$1.15\times10^{11}$\,p/b and $3.5\,\mu$m transverse emittance~\cite{LHCdesignreport}). The beam parameters achieved operationally in 2012 are summarized in Table~\ref{TAB:BeamParameters2012}.

The first part of the re-commissioning of the LHC beams in the injector chain in 2014 is focused on re-establishing the beam parameters achieved before LS1. This will rely to a large extent on the successful scrubbing of the SPS in order to suppress the electron cloud effect, which is expected to be a performance limitation during the first weeks after the start-up since large parts of the vacuum chambers have been exposed to air~\cite{Bartosik:chamonix2014}. 

\begin{table}[t]
   \centering
	\caption{Operational beam parameters in 2012.}
	\begin{tabular}{l cc }
	\toprule
	\textbf{Beam type} 	& \textbf{Intensity} 			& \textbf{Emittance} \\
	\hline
	Standard (25\,ns)	& $1.20\!\times\!10^{11}$\,p/b	& 2.6\,$\mu$m \\
	BCMS (25\,ns)		& $1.15\!\times\!10^{11}$\,p/b	& 1.4\,$\mu$m \\
	\hline	
	Standard (50\,ns)	& $1.70\!\times\!10^{11}$\,p/b	& 1.7\,$\mu$m \\
	BCMS (50\,ns)		& $1.70\!\times\!10^{11}$\,p/b	& 1.1\,$\mu$m \\
    	\bottomrule
	\end{tabular}
	\label{TAB:BeamParameters2012}
\end{table}

Once the 2012 beam parameters are reproduced, it should be possible to reach slightly higher beam intensity and potentially also higher beam brightness. Already during MDs at the end of 2012 a standard 25\,ns beam was accelerated to flat top with an intensity of about $1.3\times10^{11}$\,p/b and longitudinal beam parameters compatible with injection into LHC. In addition, high intensity LHC beams will benefit from the upgraded 1-turn delay feedback for the 10~MHz cavities and the upgraded longitudinal coupled-bunch feedback in the PS, which was commissioned in 2014. It should also be possible to enhance the beam brightness by optimising the beam production schemes as discussed at the RLIUP workshop \cite{Rumolo:RLIUP}: the space charge tune spread in the PS can be reduced by injecting bunches with larger longitudinal emittance, i.e.~increasing the bunch length and the momentum spread at PSB extraction. The maximum bunch length at the PSB-to-PS transfer is determined by the recombination kicker rise time. The maximum longitudinal emittance is determined by the RF manipulations and by the momentum acceptance at transition crossing in the PS cycle, but also by the constraint that the final bunches should not exceed 0.35~eVs for injection into the SPS. Optimising the longitudinal beam parameters at PS injection requires therefore controlled longitudinal blow-up during the PSB cycle with the C16 cavity and the use of the h=1 and h=2 PSB RF harmonics in phase at extraction to keep the larger longitudinal emittance bunches within the recombination kicker gap. Furthermore, the triple splitting in the PS was recently commissioned  at an intermediate plateau of 2.5~GeV instead of the flat bottom for providing sufficient bucket area. Further details are given in Ref.~\cite{Rumolo:RLIUP}. A summary of the expected performance limits of LHC physics beams for the run in 2015 is given in Table~\ref{TAB:BeamParameterLimitsAfterLS1}.

\begin{table}[t]
   \centering
	\caption{Expected performance limits after LS1.}
	\begin{tabular}{l cc }
	\toprule
	\textbf{Beam type} 	& \textbf{Intensity} 			& \textbf{Emittance} \\
	\hline
	Standard (25\,ns)	& $1.30\!\times\!10^{11}$\,p/b	& 2.4\,$\mu$m \\
	BCMS (25\,ns)		& $1.30\!\times\!10^{11}$\,p/b	& 1.3\,$\mu$m \\
	\hline	
	Standard (50\,ns)	& $1.70\!\times\!10^{11}$\,p/b	& 1.6\,$\mu$m \\
	BCMS (50\,ns)		& $1.70\!\times\!10^{11}$\,p/b	& 1.1\,$\mu$m \\
    	\bottomrule
	\end{tabular}
	\label{TAB:BeamParameterLimitsAfterLS1}
\end{table}

\section{SPECIAL BEAMS: DOUBLET AND 8\lowercase{b}$\oplus$4\lowercase{e}}

The doublet beam was originally proposed for enhancing the scrubbing efficiency in the SPS at low energy~\cite{IADAROLA:IPAC13}. This beam is produced by injecting a 25\,ns beam with enlarged bunch length ($\tau\!\approx\!10$\,ns full length) from the PS onto the unstable phase of the 200\,MHz RF system in the SPS.  By raising the SPS RF voltage within the first few milliseconds after injection, each bunch is captured in two neighbouring RF buckets resulting in a train of 25~ns spaced doublets, i.e.~pairs of bunches spaced by 5~ns. Very good capture efficiency (above 90\%) for intensities up to $1.7\times10^{11}$\,p/doublet could be achieved in first experimental tests in 2012.
Observations on the dynamic pressure rise in the SPS arcs confirmed the enhancement of the electron cloud activity as expected from the lower multipacting threshold compared to the standard 25\,ns beams predicted by numerical simulations~\cite{IADAROLA:IPAC13}. The experimental studies performed up to now concentrated on SPS injection energy and thus the acceleration of the doublet beam will be an important milestone during the 2014 MDs (for more details see~\cite{Bartosik:chamonix2014}).

\FloatBarrier

Thanks to its micro-batch train structure, the 8b$\oplus$4e beam was considered as an alternative to the standard 25\,ns beam in case the electron cloud remains a limitation for the operation of the LHC during the HL-LHC era \cite{Damerau:RLIUP}. Starting from 7 bunches from the PSB, the triple splitting in the PS is replaced by a direct h = 7$\rightarrow$21 bunch pair splitting, which results in pairs of bunches separated by empty buckets. Each bunch is split in four at PS flat top such that the bunch pattern 6\!$\times$\!(8b$\oplus$4e)$\oplus$8b is obtained. In this case the bunch train out of the PS is longer than the 72 bunches of the standard scheme, but the remaining gap of 4 empty buckets (about 100\,ns) is expected to be sufficiently long for the PS ejection kicker.  Without optimization of the LHC filling pattern, the total number of bunches per LHC beam is estimated as 1840. More details about the performance of this beam can be found in~\cite{Rumolo:chamonix2014}.



The estimated beam parameters are summarized in Table~\ref{TAB:BeamParameterLimits8b+4e}. Finally it should be emphasized that this beam has not been produced in the injectors so far since it was developed during LS1. First tests of this new beam production scheme will be subject of MD studies in 2014 or at latest in the beginning of 2015, depending on the availability of MD time in the injectors. 

\begin{table}[t]
   \centering
	\caption{Expected parameters of the 8b+4e beam.}
	\begin{tabular}{l cc }
	\toprule
	\textbf{Beam type} 		& \textbf{Intensity} 			& \textbf{Emittance} \\
	\hline
	Standard (8b$\oplus$4e)	& $1.80\!\times\!10^{11}$\,p/b	& 2.3\,$\mu$m \\
	BCMS (8b$\oplus$4e)	& $1.80\!\times\!10^{11}$\,p/b	& 1.4\,$\mu$m \\
    	\bottomrule
	\end{tabular}
	\label{TAB:BeamParameterLimits8b+4e}
\end{table}

\section{PROGRESS IN 2014}


The first part of the PSB and the PS startup in 2014 were devoted to the setup of the beams needed for physics. During the time of the Chamonix workshop 2014, the single bunch beam were in good shape in PSB and PS, and short trains of 12 to 24 bunches were taken in SPS for the realignment campaign and RF setting-up (energy matching). The setup of the LHC beams in the PS complex was done in parallel to physics operation and starting from re-establishing the beam conditions from 2012 (but already with the triple splitting in the PS at 2.5\,GeV instead of the flat bottom). 

The PS complex is ready to deliver the LHC beams at the startup of the SPS in September. 
As large parts of the SPS have been vented and exposed to air in the course of the works performed during LS1, it is expected that the good conditioning state of the SPS will be degraded. Therefore, two weeks of SPS scrubbing are planned for 2014 with the goal of reconditioning the SPS to the state of before LS1. The success of this scrubbing run is the critical milestone for the preparation of the 25\,ns LHC beams for physics in 2015. 

The setup of the doublet scrubbing beam for the use in the LHC will be the subject of extensive MD studies in the SPS in 2014 in several dedicated MD blocks, for establishing accelerations and  pushing the intensity to the requested $1.6\!\times\!10^{11}$\,p/doublet. During these MDs, also the behaviour of the LHC BPMs in the SPS with the doublet beam need to be tested in preparation of the LHC scrubbing, ~\cite{Lefevre:LBOC}. 

At the same time, there are many requests for dedicated MD time in the SPS for 2014 \cite{Rumolo:MSWG}. Careful planning and prioritization of studies will be crucial, as the total amount of requested dedicated MD time exceeds the MD slots available.  For example, although there are first successful recent studies in the PSB and the PS, the full qualification of 8b$\oplus$4e beam production scheme will be done in 2015. In general, it should be stressed that 2014 will be a very busy period for the injectors: Besides the physics operation after the beam commissioning with partially new or upgraded hardware, the setup and commissioning of the different LHC beams including the doublet scrubbing beam, the various dedicated and parallel MD studies, substantial amount of beam time will be needed in the PS and SPS for the first-time setup of the Ar-ion beams in preparation for the physics run beginning of 2015.

\begin{figure*}[t]
    \centering
     \includegraphics[trim= -2 0 2 0, clip,width=1\textwidth]{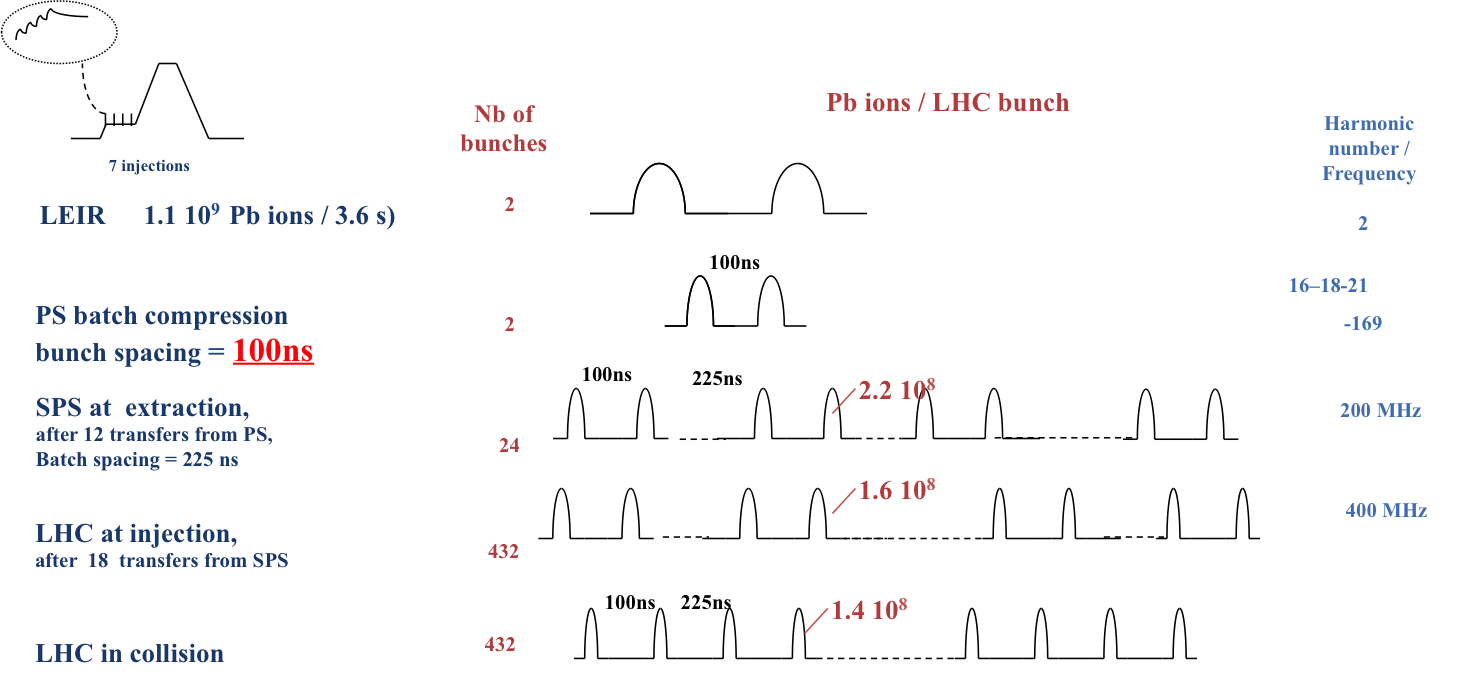}
         \caption{Ions production scheme for 2015~\cite{Manglunki:RLIUP}.}
    \label{FIG:ions}
\end{figure*}
Finally, it is worth mentioning that there will be another period of dedicated scrubbing of the SPS in 2015. While with the scrubbing run in 2014 the scrubbing efficiency and the time required for achieving acceptable conditioning after a long shutdown will be qualified, the aim of the scrubbing run in 2015 will be to condition the SPS for high intensity 25 ns beams. The outcome of these scrubbing runs will determine if the SPS vacuum chamber really need to be coated with amorphous Carbon \cite{Vallgren} as presently part of the baseline of the LIU project for suppressing the electron cloud for the future high intensity LHC beams \cite{Bartosik:RLIUP}.

\section{ION BEAMS}

The Pb-Pb run in 2011 initially projected an integrated luminosity of around 30-50~$\mu$b-1 in 4 weeks, with peak luminosity $L\text{peak} = 1.4\times10^{26}$~Hz/cm$^2$. In fact, the peak luminosity was increased to around half of the nominal ($5\times10^{26}$ Hz/cm$^2$) exceeding by far the expectations to almost 150~$\mu$b-1 integrated luminosity at 3.5~ZTeV. This was due to the increased LEIR brightness with nominal bunch population of $4.5\times10^8$ Pb$^{54+}$ ions per bunch but smaller emittances. Additional ingredients of this success were the preservation of the brightness at low energy in PS due to excellent vacuum conditions, the modified production scheme (no splitting in PS allowing half as many bunches with twice the intensity/bunch) and the good behaviour of bunches on SPS flat bottom (improved low level RF to reduce noise, IBS and space charge less critical than expected and delivered with Q20 optics after 2013). 
For the p-Pb run in 2013, the LEIR bunch intensity was further increased to$5.5\times10^8$ Pb$^{54+}$ ions per bunch, exceeding
the nominal value by a factor of 1.2. Assuming the same scheme as in 2011 and the performance of 2013, a Pb-Pb peak luminosity of $L\text{peak} = 2.3\times10^{27}$~Hz/cm$^2$ at 6.5~ZTeV can be expected. A further 20\% increase in peak luminosity can be gained by squeezing 20\% more bunches in LHC. The ion generation scheme is presented in Figure~\ref{FIG:ions} (for more details see ~\cite{Manglunki:RLIUP}). A batch compression  already tested in the PS in 2012 can allow a bunch spacing of 100~ns between two ion bunches.
Twelve of these two-bunch batches can be accumulated for every cycle of the SPS, with a batch spacing of 225~ns. After 36 injections from the SPS, assuming once again the same brightness as in February 2013, this scheme can deliver up to 432 bunches of $1.6\times10^8$ Pb$^{82+}$ ions per LHC ring, corresponding to a peak luminosity at 6.5 ZTeV of $\text{peak} = 2.8\times10^{27}$~Hz/cm$^2$.

\section{SUMMARY AND CONCLUSIONS}

Several optimizations of the beam production schemes will be implemented for the LHC Run after LS1. Single bunch beams already benefit from a better control and better reproducibility of intensity and longitudinal emittance. The longitudinal parameters at PSB-to-PS transfer of the 25\,ns and 50\,ns physics beams are optimized for allowing even higher beam brightness and, if requested by the LHC, the intensity of the 25\,ns beams can also be slightly pushed compared to the 2012 beam parameters. The first step in the beam commissioning of these LHC beams in 2014 will be however to recover their 2012 performance. In this respect, the critical milestone will be the success of the SPS Scrubbing Run, as it is expected that the good conditioning state of the SPS will be degraded due to the long period without beam operation and the venting of machine sectors related to the interventions during LS1. 

The setup of the doublet scrubbing beam with acceleration in the SPS in preparation for the LHC scrubbing in 2015 will be one of the main topics of MDs in 2014. Careful planning and prioritisation of the dedicated MDs in the SPS will be crucial due to the limited MD time available. First tests of the 8b$\oplus$4e beam already demonstrated the feasibility of the scheme and need to be tested further in 2015, in the SPS.

Besides the various physics users, the commissioning of the LHC beams and the MDs related to the new beams requested by the LHC, lots of beam time will be needed in 2014 for the first-time setup of Ar-ion beams. Regarding the ion performance, a batch compression scheme in the PS can increase the projected 2013 performance by around 20\% in peak luminosity.

\section{ACKNOWLEDGEMENTS}

The authors would like to thank G. Arduini, T. Argyropoulos, M. Bodendorfer,
T. Bohl, R. Bruce, C. Cornelis, H. Damerau, J. Esteban-MŸller, A. Findlay, R. Garoby,
S. Gilardoni, B. Goddard, S. Hancock, K. Hanke, G. Iadarola, B. Mikulec,
E. Shaposhnikova, R. Steerenberg and the PSB \& PS \& SPS Operation crews for support and helpful discussion.



\begin{thebibliography}{99}

\bibitem{Iadarola:Evian2014}
G. Iadarola and G. Rumolo, ``Electron cloud/scrubbing'', {Proceedings of the 2014 Evian Workshop on LHC Beam Operation}, CERN-ACC-2014-0319 (2014).

\bibitem{Gorini:Evian2014}
B. Gorini, ``Experiments' expectations'', , {Proceedings of the 2014 Evian Workshop on LHC Beam Operation}, CERN-ACC-2014-0319 (2014).

\bibitem{Bartosik:Evian2014}
H. Bartosik and G. Rumolo, ``Beams in the injectors'', {Proceedings of the 2014 Evian Workshop on LHC Beam Operation}, CERN-ACC-2014-0319 (2014).

\bibitem{Garoby:BCMS}
R. Garoby, ``New RF Exercises Envisaged in the CERN-PS for the Antiprotons Production Beam of the ACOL Machine'', \href{http://accelconf.web.cern.ch/accelconf/p85/PDF/PAC1985_2332.PDF}{IEEE Transactions on Nuclear Science, Vol. NS-32., No. 5} (1985).

\bibitem{Damerau:IPAC13}
H. Damerau \emph{et al.}, ``RF manipulations for higher beam brightness LHC-type beams'', \href{https://cds.cern.ch/record/1595719?ln=en}{CERN-ACC-2013-0210}.

\bibitem{Rumolo:RLIUP}
G. Rumolo \emph{et al.}, ``Expected performance in the injectors at 25 ns without and with LINAC4'', \href{https://indico.cern.ch/event/260492/}{proceedings of the Review of LHC and Injector Upgrade Plans Workshop} (2013).

\bibitem{Damerau:RLIUP}
H. Damerau \emph{et al.}, ``LIU: Exploring alternative ideas'', \href{https://indico.cern.ch/event/260492/}{proceedings of the Review of LHC and Injector Upgrade Plans Workshop} (2013).

\bibitem{Hancock:Indivs}
S. Hancock, \href{http://cds.cern.ch/record/1562030?ln=en}{CERN-ATS-Note-2013-040 MD} (2013).

\bibitem{BartosikRumolo:vdM}
H. Bartosik and G. Rumolo, \href{http://cds.cern.ch/record/1590405?ln=en}{CERN-ACC-NOTE-2013-0008} (2013).

\bibitem{Bartosik:Q20} 
H.~Bartosik {\em et al.}, ``Increasing instability thresholds in the SPS by lowering transition energy'', 
\href{http://cds.cern.ch/record/1464095?ln=en#}{CERN-ATS-2012-177}, IPAC2011.

\bibitem{Papaphilippou:Q20} 
Y.~Papaphilippou {\em et al.}, ``Operational performance of the LHC proton beams with the SPS low transition energy optics", 
\href{https://cds.cern.ch/record/1581446/files/CERN-ACC-2013-0124.pdf}{CERN-ACC-2013-0124}, IPAC2013.

\bibitem{LHCdesignreport} 
LHC Design Report, edited by O.~Br\"{u}ning, P.~Collier, P.~Lebrun, S.~Myers, R. Ostojic, 
J. Poole, P. Proudlock, \href{https://cds.cern.ch/record/782076?ln=en}{CERN-2004-003-V-1}

\bibitem{Bartosik:chamonix2014} 
H.~Bartosik {\em et al.}, ``SPS scrubbing 2014'', these proceedings.

\bibitem{IADAROLA:IPAC13}
G. Iadarola \emph{et al.}, ``Recent electron cloud studies in the SPS'', \href{https://cds.cern.ch/record/1577449?ln=en}{CERN-ACC-2013-0115}.


\bibitem{Rumolo:chamonix2014} 
H.~Rumolo {\em et al.}, ``Protons: Baseline and Alternatives, Studies Plan'', these proceedings.

\bibitem{Bartosik:LBOC}
H.Bartosik {\em et al.}, ``Special e-cloud bunch spacing: injectors'', \href{https://indico.cern.ch/event/277528/contribution/1/material/slides/1.pdf}{
LBOC meeting of November 5th (2013)}.

\bibitem{Lefevre:LBOC}
T. Lefevre et al., ``Special e-cloud bunch spacing: BI compatibility'', \href{https://indico.cern.ch/event/277528/contribution/0/material/slides/1.pdf}{LBOC meeting of November 5th (2013)}.

\bibitem{Rumolo:MSWG}
G. Rumolo and H. Bartosik, ``Injector MD planning and requests for 2014'', \href{https://indico.cern.ch/event/321789/contribution/3/material/slides/0.pptx}{MSWG meeting of May 30th (2014).}

\bibitem{Vallgren}
C. Vallgren \emph{et al.}, ``Amorphous carbon coatings for the mitigation of electron cloud in the CERN Super Proton Synchrotron'', \href{http://cds.cern.ch/record/1376711?ln=en}{Phys. Rev. ST Accel. Beams, 14, 071001} (2011).

\bibitem{Bartosik:RLIUP}
H. Bartosik {\em et al.}, ``Can we ever reach the HL-LHC requirements with the injectors?'', \href{https://indico.cern.ch/event/260492/}{proceedings of the Review of LHC and Injector Upgrade Plans Workshop} (2013). 

\bibitem{Manglunki:RLIUP}
D.Manglunki, ``Performance of the LHC ion
injectors after LS1", \href{https://indico.cern.ch/event/260492/}{proceedings of the Review of LHC and Injector Upgrade Plans Workshop} (2013). 


\end{thebibliography}
\end{document}